# The Dark Side of The Internet of Vehicles: A Survey of the State of IoV and its Security Vulnerabilities

**Tess Christensen**
1st year PhD student
Dept. of ECE
University of Florida

**Sai Bhargav Mandavilli**
1st year master's student
Dept. of ECE
University of Florida

**Chao-Yi Wu**
3rd Year PhD student
Dept. of ECE
University of Florida

## Main Report

### I. Introduction

The Internet of Things (IoT) is a technology that has become ubiquitous in the modern world. The term was first used following the creation of RFID at MIT in 1999 [1]. It has been defined as the moment in time when "more 'things or objects' were connected to the internet than people," which Cisco estimated was between 2008 and 2009 [2]. There are a multitude of connected "things" with a wide range of applications like smart homes and healthcare systems, that are categorized and illustrated in Figure 1 [4].

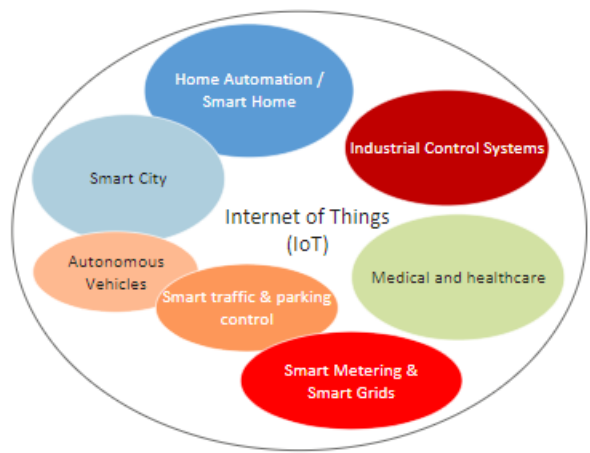

*Figure 1 - Internet of things (IoT) [4]*

One specific application, the Internet of Vehicles (IoV) intends to enhance human interactions with the world by increasing safety on the road, minimizing traffic congestion, aid in helping the environment by limiting pollution, and even help financially [18]. Intensified research of the self-driving cars was initiated by Google in 2009, the major car industries like Audi, VW, and Mercedes joined the bandwagon the following year. As automated vehicles began to hit the road for trials, development came to a screeching halt in 2018 when Uber, the popular alternative to a taxi service, was forced to suspend all self-driving testing after one of their vehicles killed a pedestrian [19]. In 2019, a Tesla driver was killed in Autopilot mode [20]. It's safe to say that automated cars are far from ready for mainstream consumers.

Certainly, the physical safety of automated vehicle drivers is top priority, however the focus of this paper takes on a different safety concern. Given that we now have just over a decade of data and research regarding IoT, it is appropriate to perform detailed analyses on its network vulnerabilities. The primary purpose of the IoT is to simplify life and create a more convenient and safer lifestyle for humans, however the large abundance of devices with minimal security can severely minimize these efforts. This paper seeks to define how cybersecurity and adversarial attacks are leveraged in IoT research, specifically with regards to the IoV.

## II. Findings

### 2.1 Security of General IoT Applications

Before a survey can be performed for what is lacking in the security of IoV, it is important to look at what research has been done on where it stemmed from, the IoT. In the past decade, a plethora of attacks have rained down on the internet of things.

*Vulnerabilities*

For hackers, IoT devices are seen as "low hanging fruit" for a few reasons such as the massive number of devices, they are perpetually connected to the internet with many flaws, and most are equipped with naïve security configurations [5]. Ellsa Bertino and Nayeem Islam [6] suggest the reasons that IoT systems are such a higher security risk than conventional computing systems are their lack of well-defined parameters that are continuously changing, their heterogeneity with respect to communication medium and protocols, platforms, and devices, the fact that the devices can be autonomously controlling other IoT devices, or even not requiring permission for installation and use. Table 1 from their paper explains the main vulnerabilities with given examples.

| Vulnerability | Examples |
|---|---|
| Insecure web/mobile/cloud interface | Inability to change default usernames and passwords; weak passwords; lack of robust password recovery mechanisms; exposed credentials; lack of account lockout; susceptibility to cross-site scripting, cross-site request forgery, and/or SQL injection |
| Insufficient authentication/authorization | Privilege escalation; lack of granular access control |
| Insecure network services | Vulnerability to denial-of-service, buffer overflow, and fuzzing attacks; network ports or services unnecessarily exposed to the Internet |
| Lack of transport encryption/integrity verification | Transmission of unencrypted data and credentials |
| Privacy concerns | Collection of unnecessary user data; exposed personal data; insufficient controls on who has access to user data; sensitive data not de-identified or anonymized; lack of data retention limits |
| Insufficient security configurability | Lack of granular permissions model; inability to separate administrators from users; weak password policies; no security logging; lack of data encryption options; no user notification of security events |
| Insecure software/firmware | Lack of secure update mechanism; update files not encrypted; update files not verified before upload; insecure update server; hardcoded credentials |
| Poor physical security | Device easy to disassemble; access to software via USB ports; removable storage media |

*Table 1 - Common IoT Vulnerabilities with examples [5]*

*Security Threats*

With such a wide range of vulnerabilities, this leads to an even wider range of adversarial attacks, with varying levels of threats, on many of the communication layers. Minhaj Ahmad Khan and Khaled Salah [9] designed a helpful taxonomy of the most common attacks, seen in Figure 2. Some common tactics of malicious actors that are typically less threatening occur at the physical and hardware layers. Low-level Sybil and spoofing attacks cause network disruptions by claiming a vast number of client identities via impersonation or falsification [7]. In addition, the physical hardware itself can be insecure by having third-party hardware components from a compromised supply chain or even from a lack of hardening, making the device easier to access, manipulate, and tamper with [8].

Many, intermediate-level, threats occur at the network layer, such as eavesdropping, denial-of-service, and fragmentation attacks. Two common fragmentation attacks on the network layer are fragment duplication attacks and buffer reservation attacks, and are more common in 6LoWPAN, the low-powered version of IPv6 that most IoT devices use. Fragment duplication attacks happen because the receiving IoT device cannot verify at the 6LoWPAN layer if a fragment originates from the same source as previously received fragments of the same IPv6, thus processing all fragments that *appear* to belong to the same IPv6 packet [3]. A buffer reservation attack exploits the scarce memory of resource-constrained nodes, and forces fragmented packets to be dropped if the reassembly buffer is occupied [3]. At the link level, which is less common, a sleep deprivation attack can occur in which the energy and lifetime of the device nodes are drained by an intruder by maximizing their power consumption [6].

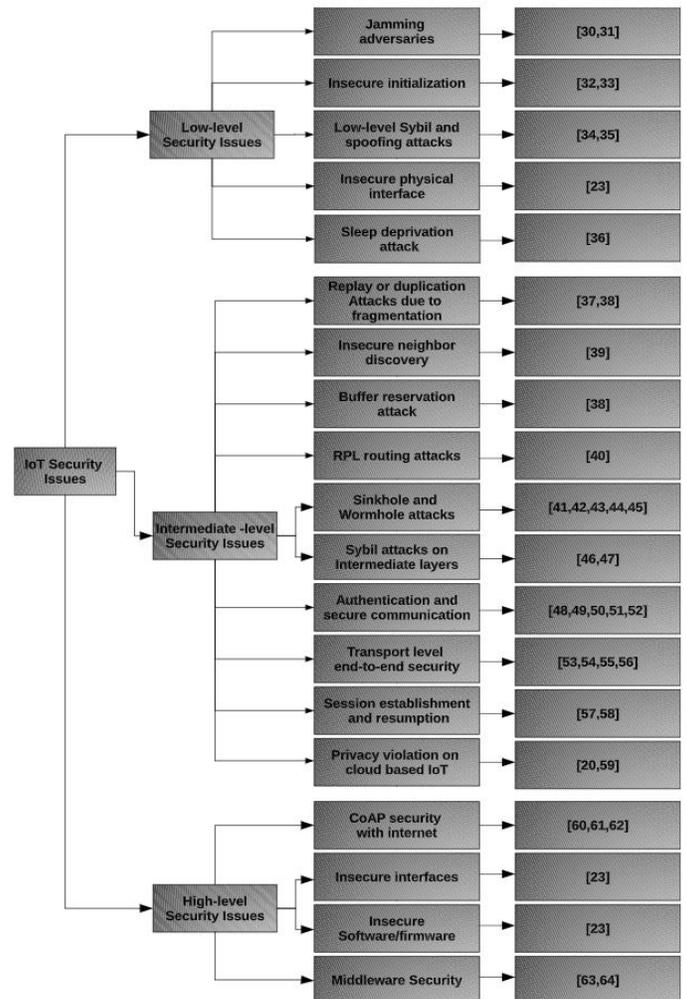

*Figure 2 - taxonomy of common security threats [9]*

*Botnets*

Though many threats were categorized in the aforementioned taxonomy, one particular threat had minimal discussion but poses a threat to many IoT systems: botnets. A botnet is defined as an automated network, or robot network, of compromised machines commonly referred to as "zombies" that are all under the control of a central device called the "botmaster" [9]. It is reported that this particular attack in the IoT doesn't receive the attention it deserves, and the fear is that more advanced forms of botnets are soon to come to possibly disrupt the Internet infrastructure itself if the security community doesn't act quickly [5]. It is important to note that not all botnets are malicious. Some are simply used for advertising or even performing research, however, the botnets discussed in this paper are of the malicious sort.

The botnet is a key part of a Distributed Denial of Service (DDoS) attack, wherein an adversary will use a botnet to make simultaneous requests to one or more servers to congest and overwhelm said servers. When the servers are overwhelmed (via bandwidth or the victim's resources), they will ignore legitimate requests from end-users an, use the attack potentially for financial gains, revenge, extortion, activism, or even simply for boasting privileges [4]. With the development and mass production of low-security IoTs in recent years, this is a new and enticing avenue for DDoS attacks, beginning with the Mirai botnet[1]. The Mirai malware was released in 2016 on by hitting a security consultant's website with 620 Gbps of traffic, and soon after targeting a French webhost and cloud service provider with 1.1 Tbps of traffic [5]. Many sites like PayPal, Netflix, and Spotify were affected, thus making this the largest DDoS attack in history.

The Mirai malware attacks internet connected IoT devices running a Linus operating system that have open telnet ports. The malicious program finds the open telnet ports by scanning IPs on port 23, then attempts to gain a Linux shell, then fingerprints the device, and finally begin scanning for and infecting other vulnerable devices. Kolias, et al. [9] details the four main components of the botnet: the *bot* malware that infects devices, the *command and control (C&C) server* that acts as the centralized management interface, the *loader* that disseminates executables, and the *report server* that maintains the database containing details on the zombie device. A diagram detailing the key steps in botnet operation and communication using these primary components is shown in Figure 3.

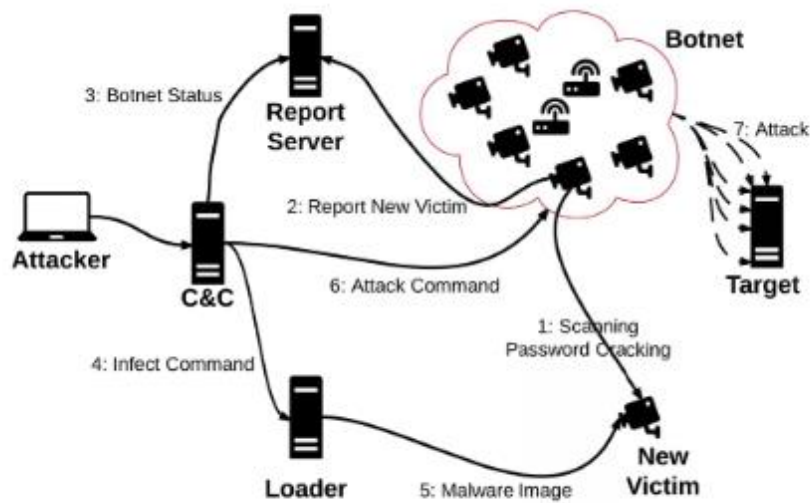

*Figure 3 - Overview of the Mirai communication with its main steps and components. [10]*

---

[1] Mirai is Japanese for "the Future" [9]

Shortly after the 2016 attack, the Mirai creator released the source code of the malware, leading to the assumption that better defense and detection mechanisms, but it only led to more variants of the malicious software, like the Hajime botnet just one month later [9]. As the suggestion that higher-powered and larger smart devices like autonomous cars will be a reality, it is concerning that not more research has been performed on this topic.

2.2 Automated Vehicles

In this section, two technologies are introduced to realize the smart vehicular network (SVN). The first technology is discussed from the work [11] which proposes a cooperative scheme to improve the capacity of vehicular networks, and the second covers the work [12] which proposes a methodology to operate the vehicular communication in the millimeter-wave band is discussed and analyzed. It is noted that both technologies can improve the channel capacity, and hence can support the SVN. With SVN, enhanced road traffic efficiency, improved road safety, and realize the autonomous driving are valuable benefits.

*Cooperative vehicular scheme*

The authors in [11] propose a cooperative communication scheme which takes the vehicle-to-infrastructure (V2I) communications, the vehicle-to-vehicle (V2V) communications, and the density of the mobile vehicles. They also provide a capacity analysis for the proposed scheme. To further discuss the proposed scheme, it is important to introduce the network model and the wireless communication model for V2I and V2V.

For the network model, the authors consider a bi-directional highway segment with length L and assume that the roadside infrastructure (base stations) are uniformly deployed on the highway. Additionally, it is also assumed that the traffic model follows a Poisson process with densities ρ1 and ρ1, respectively. For the wireless communication model, the authors assume that all the base stations and the vehicles have the same radio range. Note that the base station (or vehicle) can communicate with the vehicle if the Euclidean distance is smaller than the radio range. The half-duplex radio is adopted in this work, which means that the antenna on the vehicle can only transmit or receive the signal at the same time. Moreover, the unicast scenario is considered. That is, each base station or vehicle can only send the information to one vehicle at a time. To proceed, the authors assume that different channels are allocated for the V2I and V2V communications. Hence, there is no mutual interference among the V2I and V2V communications. Finally, the CSMA media access control (MAC) protocol is adopted for the V2V communications. With these preparations, the cooperative communications scheme proposed in [11] is described in the following paragraph.

Suppose that certain vehicles want to obtain information (files) from the remote server. Then each requested file will be divided into multiple small pieces and sent to different base station. By doing so, each base station can have different portions of the file which allows the cooperation among the base stations. Moreover, the delivered data can be further split and send to the vehicles which require this information. The data will be also sent to the helpers. Here, the helpers are those vehicles which do not require this information but can receive the data since they move into the coverage range of the base station. Notice that each helper will store the data for different vehicles and is willing to send this information. Besides, the authors assume that there exists a central server which has the full knowledge of the network topology and the transmitted data. This assumption indicates that the data received by the helper is the information that the target vehicle wants. With this cooperative scheme (setting), the vehicles can receive the required data when they move into the coverage of base stations. The vehicles can also receive the required data from the helpers when they are outside of the base station. Thus, the proposed scheme can increase the capacity.

To define the capacity proposed in [11], consider the time interval [0,t] and the amount of data $D_x(t)$ being transmitted in this time interval. Note that $D_x(t)$ includes the data obtained directly from the base station and indirectly from the helpers. Let φ be the set of all the scheduling algorithms. Then the maximum average data rate (γ) is given by

$$\gamma = \min_{x \in \phi} \gamma_x = \min_{x \in \phi} \lim_{t \to \infty} D_x(t)/t. \quad (1)$$

Note that the achievable rate in (1) includes the capacity achieved by the V2I communications and the capacity achieved by the V2V communications. The authors in [11] provide a detail analysis for (1) and show via simulations that the proposed cooperative scheme can achieve excellent rate. Hence, the proposed scheme can improve the capacity of the vehicular networks.

*Millimeter-wave communications*
Millimeter-wave (mmWave) communication has become an outstanding technology for autonomous driving since it has large bandwidth to provide high data rate. However, realizing the mmWave communication faces many challenges. For example, directional communication between the base station and the vehicle requires accurate beam alignment. The performance of mmWave systems will be degraded if the beam alignment is inaccurate. Moreover, due to the high penetration loss, the mmWave signals are prone to blockages (See Figure 4). To overcome this problem, one possible solution is to allow the base station to perform beam selection based on the surrounding environments. In work [12], the authors approach the beam selection problem by modeling it as a contextual multi-armed bandit problem. The provided model can be extended to different contents. The authors also provide an online learning algorithm to solve the multi-armed bandit problem. Via simulations, the authors show that the proposed algorithm is effective and can work well in many environments. In the following, the system model considered in this paper and the online learning algorithm proposed in [12] are introduced.

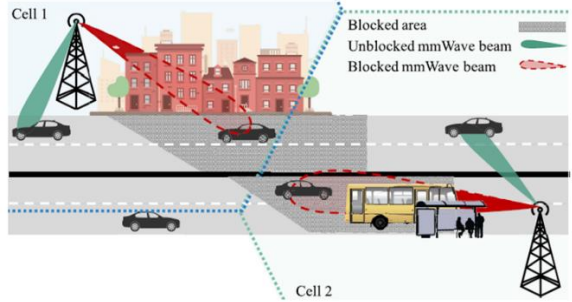

*Figure 4 - The impacts of blockage for mmWave systems. This figure comes from reference [12].*

The work [12] considers a cellular system where mmWave base stations (mmBSs) provide services for the vehicles. The downlink scenario is considered in this work. Note that each vehicle is equipped with an LTE interface to make connections with the base station. Each vehicle also has the mmWave interface to keep high-speed communications. To formulate the beam selection as a contextual multi-armed bandit problem, the authors denote rb(x) as the beam performance. Note that rb(x) is also a random variable which indicates the amount of data being successfully received from the base station using the beam b. With the previous definition, the beam selection problem can be formulated by choosing the beam b* which maximizes R(T). Here, R(T) is given by

$$R(T) = \sum_{t=1}^{T} \sum_{i=1}^{V_t} \sum_{j=1}^{m} \left( \mu_{b_{t,j}^*(\chi_t)}(x_{t,i}) - E[\mu_{s_{t,j}(\chi_t)}(x_{t,i})] \right), \quad (2)$$

Where T is a finite time horizon, $V_t$ is the number of vehicles registered for the base station, m is the number of beams, and $\mu_b$ is the expected beam performance of beam b under context x.

To solve the aforementioned problem, the authors propose an online learning algorithm to deal with it. Note that this approach is reasonable as the online learning architecture can take the dynamic traffic into consideration. Specifically, the proposed algorithm first partition the contexts into small sets. Then from each of the sets, the algorithm learns the performance of different beams independently. By doing so, the proposed algorithm can select the best beam to serve the chosen vehicles with the desired context. Simulation results show that the proposed online learning algorithm can achieve excellent performance.

## 2.2 IoV Security

Security in IoV becomes of paramount importance as any system failure directly affects the safety of the driver. Mentioned below are some of the common attacks and threats to security in the IoV environment.

Denial of Service (DoS) attacks: DoS make substantial exercises of a framework inaccessible. Due to this, components (or ECUs) in the vehicle cannot communicate with each other and the central system will not get organized data. For example, radar systems for lane automation bring unexpected outcomes. In a Distributed Denial-of-Service (DDoS) attack, nodes can attack the IoV system from various areas making its discovery harder. This attack can hurt the vehicles in the system by completely disconnecting several components of the network.

Impersonation attack: Here, malicious attackers pretend to be legitimate users and send reports to earn benefits and thus insert forged signatures and reports which can confuse and mislead customers. In an IoV environment, attackers can send malicious data related to navigation, vehicle performance etc.

### Sybil attack

One of the most popular attacks on vehicular ad-hoc networks (VANET) [13] in which the attacker insidiously claims or takes different symbols and utilizes their traits to compromise the usefulness of VANET by diffusing false characters. To undergo a Sybil attack, the attacking nodes must possess different sets of authentication certificates which can be obtained while capturing or spoofing from the vehicles while they handshake with the Road Side Units (RSU) and use these fake identities to obtain the required Temporary Certificate (TC) and communicate like an authentic vehicle. These nodes further form as attacking nodes covered under different RSUs and can send false information to nearby vehicles to send false locations at a given period.

### Replay attack

This form of threat exploits the conditions of the network by storing the messages to reuse it later when the data becomes invalid.

### Eavesdropping attack

Also called as the network packet sniffing or snooping attack. Here, an unauthorized party takes control of the communication between the internet and the vehicle network to modify and erase system data that is exchanged on the bus (here the control information between ECUs).

Table 2 illustrates different security and privacy issues and draws a correlation to express all the issues familiar to IoV networks. This data helps identify serious issues and thus causes us to act on them to infer a solution.

|  | [13] | [15] | [26] | [22] | [23] | [24] | [12] | [21] | [25] | [17] | [19] | [16] |
|---|---|---|---|---|---|---|---|---|---|---|---|---|
| Trusted Third-Party Validation | ✓ | ✓ | ✓ | ✓ | ✓ | ✓ | ✓ | ✓ | ✓ | ✓ | ✓ | ✓ |
| Data encryption |  | ✓ | ✓ | ✓ | ✓ | ✓ | ✓ | ✓ | ✓ | ✓ | ✓ | ✓ |
| Authentication | ✓ | ✓ | ✓ | ✓ | ✓ | ✓ | ✓ | ✓ | ✓ | ✓ | ✓ |  |
| Sybil Attacks | ✓ |  |  |  |  |  | ✓ |  |  |  |  |  |
| Impersonation Attacks | ✓ |  | ✓ | ✓ |  |  | ✓ | ✓ |  |  |  |  |
| Privacy Techniques for vehicles | ✓ | ✓ |  |  | ✓ | ✓ |  | ✓ | ✓ |  | ✓ |  |
| Privacy techniques for navigation | ✓ | ✓ | ✓ | ✓ | ✓ | ✓ |  | ✓ |  |  |  |  |
| Fairness | ✓ | ✓ | ✓ | ✓ | ✓ |  |  |  |  |  |  |  |
| DoS Attacks |  |  | ✓ | ✓ |  |  |  |  |  | ✓ |  |  |
| DDoS Attacks |  |  |  |  |  |  | ✓ |  |  |  |  |  |
| Data non-repudiation | ✓ | ✓ | ✓ | ✓ | ✓ | ✓ |  |  |  |  |  |  |
| Access Control | ✓ | ✓ |  | ✓ | ✓ | ✓ |  |  |  |  |  |  |
| Replay Attacks |  |  | ✓ | ✓ |  |  |  | ✓ |  | ✓ |  |  |
| Forgery Attacks | ✓ |  | ✓ |  |  |  |  |  |  | ✓ |  |  |
| Eavesdropping attack |  |  |  |  |  |  |  |  |  | ✓ |  |  |
| Auxiliary attack |  |  |  |  |  | ✓ |  |  |  |  |  |  |
| Collusion Resistance |  | ✓ | ✓ |  |  | ✓ | ✓ |  |  |  |  |  |
| Data Integrity | ✓ | ✓ | ✓ | ✓ | ✓ | ✓ |  | ✓ | ✓ |  |  |  |
| Man in the Middle Attacks |  |  | ✓ |  |  |  |  |  |  |  |  |  |
| Session Key Security |  |  | ✓ |  |  |  |  |  |  | ✓ |  | ✓ |

*Table 2- Privacy and security issues addressed in previous literature*

Major problems as described above act as a barrier for a safe and secure network that hinders the privacy of the vehicle data and its users. Following are a few solutions to enhance the quality of the architecture and prevent mishaps due to attacks, a) authentication guarantees that any components involved in transmission can verify their authenticity. Sensitive data in the vehicle network framework can be encrypted. However, this method triggers a huge overhead in terms of information sharing and searching in IoV, for example, data is shared between vehicles to analyze real-time information of traffic on the road, b) access control intends to legalize elements added to the network. With the help of ECUs, control can be given to trusted third party who authenticates the element before adding or giving information about the On-Board Units (OBUs), c) non-repudiation guarantees that any article in the framework cannot involve in gathering or distributing information from the network.

Additionally, integrity in a vehicular computing framework guarantees that the trustworthiness pre-requisite is satisfied since unsanctioned alteration of transmitting data may bring about genuine as well as appalling outcomes, particularly in critical vehicular applications. The service provider for network connectivity must ensure that the fog nodes or cloud servers should consistently be prepared for recovery or transmission of information figures.

Reliability guarantees that data collected from smart vehicles are not fictitious or untrue. For this, the routers must propagate the data to the central repository based on geographical location, preventing confusion of data [14]. Data received must also be flagged with a timestamp and taken into consideration for only a particular time slot after which the packet will be discarded and new information will have to be resent from the vehicle. Also, the vehicle sending the data must be registered with the system. Trust is an important concept to accomplish security and safety within the system, where each vehicle established in a trustful model can assess the dependability of the received data and its sender.

Similarly, an eavesdropping attack accesses the location of the vehicle, its driving information and to prevent this type of attack, pseudonyms can be deployed which changes the name of the vehicle and the attackers would find it difficult to track the activities of the vehicle, thereby maintaining the individual privacy. Unauthorized user data should not be injected into the system as it can severely hamper the performance of the network [15].

Anonymity techniques such as group signatures and k-anonymity can be used to protect the vehicle's privacy. These techniques prevent attackers from separating the vehicles dependent on swarm detecting information. Whether or not swarm detecting information is conveyed with a signature, vehicles might need to inform their group detecting status without unveiling their characters. This activity may present an opportunity for the attacker to access some fragile (critical) data. Subsequently, how to design the cloud to conduct these group detecting assignments is fundamental for the vehicles with respect to security. Another technique to achieve security is by actualizing outrageous focuses protection [16] which abuses the idea of end system focuses that are regular to vehicle social networks (VSNs) to make shared zones to anonymize them.

*Focus topic:* Fighting Sybil attacks in IoV

Figure 4 depicts a vehicle V that enters an RSU coverage from another RSU, it must submit a valid certificate before establishing communication with other vehicles connected to the same RSU.

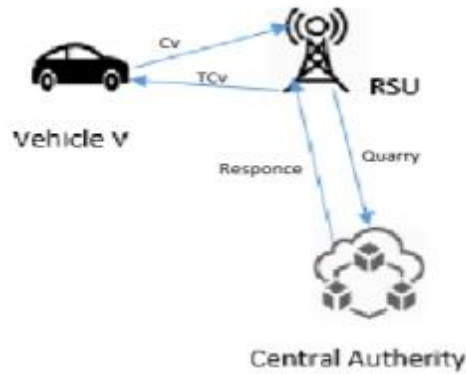

*Figure 4- Communication model of vehicle, RSU and central authentication system*

All RSUs relate to one another to a secured central repository equipped with big data analytics. They may transfer information such as certificates, GPS co-ordinates and may conduct necessary computation with a Quarry and Central Authority system. The central authority keeps track of valid certificates and maintains revocation lists and it also maintains a blacklisted vehicles database and updates all its information on a timely basis.

To undergo a Sybil attack, the attacking nodes must possess different sets of valid authorization certificates, which may be acquired while capturing or spoofing from the vehicles while they communicate with the RSUs and use these fake certificates to gain the required TCv from the RSU. The nodes then talk to the vehicles and at any given moment of time, they may be registered under different RSUs with different certificates. From the set of vehicles, few of the nodes may behave as attacking nodes under different RSUs and can send false information such as false location to nearby vehicles.

To overcome this attack, every vehicle sends location information to the RSU at regular intervals of time. Here, the K-Means Clustering algorithm as proposed in [17] can be utilized to form vehicle clusters under each RSU which leads in different clusters. Due to the dynamic nature of the vehicles, the clusters may change dynamically thereby resulting in formation of new clusters due to a change in location.

In the algorithm, mean value for each cluster is calculated for n vehicles under each RSU for every 0.5 to 2 seconds in new set of means and these values are recorded. To detect a suspicious node, if two consecutive mean variations are more than the specified threshold, study the nodes that caused these variations and mark them as suspicious nodes. For every node in this list, if speed variation is much high, remove it from the suspicious set.

The next step is to filter out the Sybil nodes. To do this, once after getting M = K-Means of location data, i.e., $M_{t1}$, $M_{t2}$... $M_{tx}$, calculate the mean variation which can be drawn from the difference in two consecutive means,

$$\delta M_1 = M_{t2} - M_{t1}$$

$$\delta M_2 = M_{t3} - M_{t2}$$

$$\delta M_3 = M_{t4} - M_{t3}$$

Let λ be the Mean Variation Threshold which can be derived from the previous outcomes, type of the road, speed limitation of the vehicles etc. The algorithm then sets value of λ considering all these factors based on the conditions of the road. Once the threshold is fixed, then it is compared against the mean change, as follows,

$$\delta M_1 > \lambda, \delta M_2 > \lambda$$

Then, the nodes which caused this variation are identified into a set called the *suspicious nodes set*.

## III. Conclusion

For the smart vehicular network, we studied two technologies to realize it. The first technology is the cooperative scheme which improves capacity by properly combining the V2V and V2I. The second technology is an online learning algorithm which can deal with the beam selection problem in mmWave system. Both are effective and can be used in autonomous driving systems. However, advancements in the field of IoV have elicited research in different areas related to the field. This highlights a critical need to address security and protection challenges as a result of the progression of vehicles and everything that is being transferred to the internet. In addition, to understand exactly where research is missing regarding IoV, we found that a survey of current research in the vulnerabilities and threats to general IoT applications. In addition to other attacks, we found that DDoS attacks in the form of botnets are significant threats to the IoT world. Upon researching which threats and vulnerabilities are leveraged in IoV research, the field was severely lacking in botnet and DDoS attack research. If developers neglect to address this issue before interconnected vehicles become a mainstream reality, this discovery can have severe ramifications for the safety of IoV consumers around the globe.